\documentstyle[epsfig,aps]{revtex}
\begin{document}
\draft
\title{r-modes in the Tolman VII solution}
\author{Nicholas Neary\cite{email} and Kayll Lake\cite{email1}}
\address{Department of Physics, Queen's University, Kingston, Ontario, Canada, K7L 3N6 }
\date{\today}
\maketitle
\begin{abstract}
The r-mode frequencies of the Tolman VII solution for the slowly
rotating non-barotropic approximation within the low frequency
regime are estimated. The relativistic correction to Newtonian
r-mode calculations is shown as function of the tenuity
$\frac{R}{M}$ and is shown to be significant only for very compact
neutron stars.
\end{abstract}
\section{Introduction}
The possibility that the fluid motion of rotating neutron stars is unstable to the emission
of gravitational radiation has motivated numerous recent studies of the so-called ``r-mode" instability \cite{Rev}.
Although this is a relativistic effect, arising from perturbations of the spacetime, the majority of available work
has been focused on Newtonian and post-Newtonian approximations.  In the first-order (in the angular velocity
$\Omega$) relativistic treatment the eigenfrequencies are real and the modes are determined by one ordinary
differential equation\cite{Kojima} in the low frequency regime for non-barotropic \cite{Bar} stars.  This
singular \cite{Kok1} eigenvalue problem then gives rise to a continuous spectrum, but can also admit a
discrete r-mode solution\cite{Lock}(At higher frequencies the r-modes exhibit complex eigenfrequencies,
losing their singular structure.) It has been further shown \cite{Kok2} that if the mode occurs inside the star,
then it is associated with a diverging velocity perturbation. It has been shown that the r-modes of low-index
realistic polytropic models exhibit this unphysical feature, while constant density  models do not\cite{Kok2},
suggesting that the r-mode instability might be a curiosity of unphysical models.  It is more likely, however,
that it is the mathematical description of the problem that is failing, not the underlying physics.   Taking
this view, we present a straightforward way to estimate the relativistic correction to Newtonian r-mode
frequencies for realistic equations of state.  Calculations are then peformed in the slowly rotating
counterpart to the Tolman VII solution - the exact solution of Einstein's field equations that is currently
the best fit to a neutron star \cite{Lat}.  We show that the maximum relativistic correction to Newtonian
r-mode calculations based on this model is $< \sim 23\%$ for entirely causal solutions.
\section{The Eigenvalue Problem}
The static spherically symmetric metric in "curvature" coordinates \cite{Units} is
\begin{equation}
ds^2 = ds^2_{\Gamma} + r^2d\theta^2+r^2sin^2(\theta)d\phi^2, \label{static}
\end{equation}
where $d^2_{\Gamma}$ is a static Lorentzian 2-surface with coordinates $(t,r)$. The source of
(\ref{static}) is considered to be a perfect mathematical fluid (of isotropic pressure $p(r)$
and energy density $\rho(r)$) generated by radial streamlines of constant $r$.
To first order in the angular velocity ($\Omega$), a rotating star can be described by the
stationary axisymmetric metric
\begin{equation}
ds^2=-e^{\nu(r)}dt^2+e^{\lambda(r)}dr^2+r^2d\theta^2+r^2sin^2(\theta)_d\phi^2-2w(r)r^2sin^2(\theta)dtd\phi,
\end{equation}
where the functions $\nu(r)$ and $\lambda(r)$ are identical to those in (\ref{static}). The function
$w(r)$ satisfies Hartle's equation \cite{Hartle}
\begin{equation}
\frac{1}{r^4}\frac{d}{dr}(r^4j\frac{d\overline{w}(r)}{dr})+\frac{4}{r}\frac{dj}{dr}\overline{w}(r)=0,
\end{equation}
subject to the boundary condition
\begin{equation}
1=\Omega=[\overline{w}(r)+\frac{R\frac{d\overline{w}(r)}{dr}}{3}]_{r=R},
\end{equation}
where $j=e^{-\frac{\nu(r)+\lambda(r)}{2}}$ and $\overline{w}(r)=\Omega-w(r)$. The axial modes for
non-barotropic stars are determined by Kojima's equation \cite{Kojima}
\begin{equation}
 (\alpha-\overline{w}(r)) ({e^{-\lambda(r)}}h_o^{''}-4\, r\pi\Upsilon(r)h_o^{'}- (8\,\pi \, \Upsilon(r)-4\,{\frac {M}{{r}^{3}}}+{\frac {l (l+1)}{{r}^{2}}})h_o
)+16\,\alpha\,\pi \, \Upsilon(r)h_o + O(\Omega^2) + ... =0, \label{Kojima}
\end{equation}
where $^{'} \equiv \frac{d}{dr}$,
\begin{equation}
\Upsilon(r)=p(r)+\rho(r),
\end{equation}
and $h_o$ is a perturbation variable ($l$ and $m$ are spherical
harmonic indices).  Note that we have included higher-oreder
corrections to this description.  The problem is singluar at $r_s$
when $\overline{w}(r_s)=\alpha$, the eigenvalue, which is related
to the mode frequency ($\sigma$) by
\begin{equation}
\sigma=-m\Omega(1-\frac{2\alpha}{l(l+1)}).
\end{equation}
Newtonian r-modes have $\alpha=1$.  The fluid velocity perturbation ($u_s$) vanishes at the boundary defined
by the vanishing of the isotropic pressure. It is related to the perturbation variable by
\begin{equation}
u_s=\frac{2m\overline{w}(r)}{2m\overline{w}(r)-l(l+1)(\sigma+m\Omega)}e^{\frac{\lambda(r)-\nu(r)}{2}}h_o.
\end{equation}
Kojima's equation (\ref{Kojima}) admits a continuous spectrum from the centre ($\overline{w}(0)$) to the
boundary ($\overline{w}(R)$) \cite{Kok1}, and may also admit discrete r-mode solutions \cite{Balmforth}.
The singular point corresponding to these solutions cannot be inside the star, as the velocity perturbation
would diverge \cite{Kok2}.  In this scenario, therefore, eigenvalue solutions are only physical if
$\overline{w}(R)\leq\alpha\leq1$, whereas many realistic equations of state yield  $\alpha<\overline{w}(R)$.
It is likely that this unphysical behaviour arises from the neglect of higher order terms near the r-mode,
since the r-mode instability appears even in Newtonian theory \cite{Lock2}.  These higher-order terms can
only be significant when the $16\,\alpha\,\pi \, \Upsilon(r)h_o$ term in (5) is small.  For realistic
equations of state, $\rho(r)$ monotonically decreases through the solution, reaching a minimum at the
boundary, which is defined by $p(R)=0$.  Hence, the eigenvalues must correspond to a value of $\overline{w}(r)$
close to the boundary of the solution.  Therefore a useful approximation for computing the relativistic
correction to Newtonian r-mode calculations for realistic equations of state is that $\alpha\approx\overline{w}(R)$.
\section{The Tolman VII Solution}
The Tolman VII solution \cite{Tolman} is one of only two \cite{Buch} known physically reasonable exact
spherically symmetric perfect fluid solutions of Einstein's equations where the energy density vanishes at
the boundary \cite{Del}.  The associated metric is
\begin{equation}
ds^2_{\Gamma}= -\sin(\ln (\sqrt {(\sqrt {1-5\,{\frac {{\xi}^{2}}{\gamma}}+3\,{\frac {{\xi}^{4}}{\gamma}}}+{\frac {{\xi}{2}\sqrt {3}}{\sqrt {\gamma}}}-5/6\,{\frac {\sqrt {3}}{\sqrt {\gamma}}}){C}^{-1}}))^2dt^2+{R}^{2} (1-5\,{\frac {{\xi}^{2}}{\gamma}}+3\,{\frac {{\xi}^{4}}{\gamma}})^{-1}dr^2
\end{equation}
where $C$ is a number chosen to give zero energy density at the boundary \cite{Neary} and is given by
\begin{equation}
C=\frac{2}{\sqrt{3\gamma}}\sqrt{\frac{12\,{\gamma}^{3/2}-23\,\sqrt {\gamma}+4\,\sqrt {3}\sqrt {\gamma\,
 (\gamma-2)}}{e^{4\arctan(\varsigma)}}},
\end{equation}
where
\begin{equation}
\varsigma={\frac {3\,{\gamma}^{3/2}-6\,\sqrt {\gamma}+\tan(\ln (2))\sqrt {3}
\sqrt {\gamma\, (\gamma-2)}}{-3\,{\gamma}^{3/2}\tan(\ln (2
))+6\,\sqrt {\gamma}\tan(\ln (2))+\sqrt {3}\sqrt {\gamma\, (
\gamma-2)}}},
\end{equation}
$\xi=\frac{r}{R}$, and $\gamma$ is the tenuity (total radius to mass ratio, $\frac{R}{M}$).  The solution
is causal \cite{causal} for $\gamma>\sim3.707$.  The isotropic pressure is not a ``elementary" function.
In contrast, the energy density is simply
\begin{equation}
\rho(\xi)=\rho_c(1-\xi^2),
\end{equation}
where $\rho_c=\frac{15}{8 \pi R^2 \gamma}$.  Despite its simplicity, this density profile (and the
corresponding binding energy and moment of inertia determined from it) is in good agreement with
realistic equations of state for neutron stars with $M>1.2M_\odot$ \cite{Lat}, even in the star's
intermediate regions.  Another appealing feature of this solution is that it does not have a phase
transition (in $\rho$) at the boundary.  Since the Tolman VII solution can reproduce several features
that are independant of the neutron star equation of state, we take the view that using it is the best
way by which to gauge the relativistic correction to Newtonian r-mode calculations.
\section{Results and Conclusions}
Numerical integrations of $(3)$ in the Tolman VII solution give
the plot shown in Figure 1. At the causal limit (subluminal
adiabatic sound speed throughout the interior) the eigenvalue is
$\alpha\approx\overline{w}(R)\approx0.77$ which gives an estimate
of $23\%$ for the correction to the Newtonian r-mode calculation.
The size of the correction then decreases monotonically with
increasing $\gamma$. We find, therefore, that the difference
between the r-mode frequencies of realistic slowing rotating
non-barotropic neutron stars and those of Newtonian models in the
low-frequency approximation, is only significant for very compact
neutron stars.
\section*{Acknowledgments}
This work was supported by a grant (to KL) from the Natural
Sciences and Engineering Research Council of Canada.    We thank
Fred Nastos for comments pertaining to the construction of the
figure. We would also like to thank Don Page for several very
helpful comments on the Tolman VII solution.  The comments of a
reviewer helped to significantly improve the presentation of the
ideas in this paper. Portions of this work were made possible by
use of \textit{GRTensorII}\cite{grt}.

\newpage
\begin{figure}
\center{\psfig{file=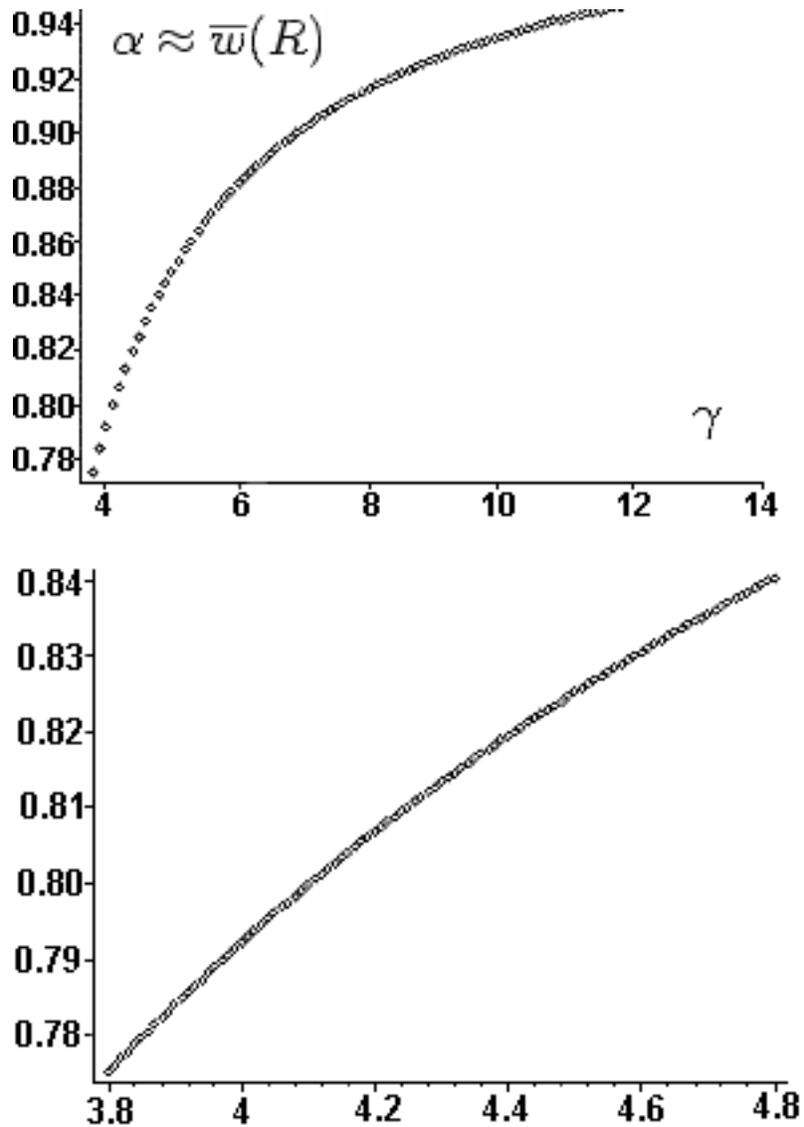}}\\ \caption{ Numerical integration of
$(3)$ in the Tolman VII solution. The solution is entirely causal
for $\gamma> \sim 3.707$. An enlarged view of the compact region
is shown. At the causal limit (subluminal adiabatic sound speed
throughout the interior) the eigenvalue is
$\alpha\approx\overline{w}(R)\approx0.77$ which gives an estimate
of $23\%$ for the maximum correction to the Newtonian r-mode
calculation. Note that the size of the correction then decreases
monotonically with increasing $\gamma$.} \label{fig1}
\end{figure}


\begin{references}
\bibitem{Bar}Following \cite{Kok2} we call a stellar model barotropic if the unperturbed
configuration obeys the same equation of state as the perturbed configuration.
\bibitem{Units} We use geometrical units throughout. For example, the tenuity $R/M$ is actually $c^2R/GM$.
\bibitem{causal} By causal we mean a subluminal adiabatic sound speed ($\textrm{v}_s^2 = \frac{dp}{d\rho}
< 1$).
\end{references}
\end{document}